\documentclass[conference,letterpaper]{IEEEtran}

\usepackage{cite}
\usepackage[utf8]{inputenc} 
\usepackage[T1]{fontenc}
\usepackage{url}
\usepackage{ifthen}
\usepackage{cite,enumitem}
\usepackage[cmex10]{amsmath} 
\usepackage{graphicx,mathtools}
\usepackage{amssymb,amsthm,verbatim}
\usepackage[colorlinks=true,citecolor=blue]{hyperref}
\usepackage[most]{tcolorbox} 
\usepackage{mathrsfs}  
\usepackage{dsfont}
\usepackage{booktabs}
\usepackage{subcaption}
\tcbuselibrary{listingsutf8} 

\definecolor{lightgray}{RGB}{224,224,224}

\newtheorem{example}{Example}

\newtheorem{remark}{Remark}

\newcommand{\Oh}{\mathcal{O}}

\newcounter{myboxctr}
\newtcolorbox{myrefbox}[2][]{%
  colback=gray!10,
  colframe=black!55,
  fonttitle=\bfseries,
  before upper={},
  title={\refstepcounter{myboxctr}\label{#1}Box~\themyboxctr:~#2}
}
\begin{document}
\title{Universal Maximum Likelihood (List) Decoding via Fast Vector-Matrix Multiplication}

\author{%
  \IEEEauthorblockN{Hoang Ly, Michael Schleppy, and Emina Soljanin}
  \IEEEauthorblockA{
  Department of Electrical \& Computer Engineering, Rutgers University}
                    E-mail: \{\texttt{mh.ly;michael.schleppy;emina.soljanin}\}@rutgers.edu
}
\maketitle
\begin{abstract}
Maximum-likelihood (ML) decoding for arbitrary block codes remains fundamentally hard. Its computational complexity—measured by the total number of multiplications—is no better than that of exhaustive search, which requires \( q^k n \) operations for an \([n, k]_q\) code. This paper introduces a simple, code-agnostic framework that, for fixed-rate codes (i.e., when \(k/n\) remains constant), reduces the complexity of ML decoding by a factor of \(n\), achieving \(q^k\) operations—a highly practical and desirable improvement. The result holds for both linear and nonlinear block codes over general memoryless channels and under both hard-decision and soft-decision decoding. It naturally extends to intersymbol-interference (ISI) channels and list decoding with only a negligible increase in complexity. Our core insight is that, upon receipt of each sequence at the receiver, the conditional probability of that sequence for each codeword in the codebook (i.e., the \emph{likelihood}) can be expressed as the inner product of two carefully constructed vectors—the first depending on the received sequence, and the second on that codeword itself. As a result, evaluating the likelihoods for all codewords in the codebook reduces to a single vector-matrix multiplication, and ML decoding (MLD) becomes the simple task of picking the maximum entry in the resulting vector. The only non-trivial cost lies in the vector-matrix product. However, our construction allows the use of the Mailman algorithm to reduce this cost. This time reduction is achieved at the cost of higher space complexity, requiring $\mathcal{O}(q^{k})$ space to store the pre-computed matrix containing information of the entire codebook.

\end{abstract}
\section{Introduction}\label{sec:intro}
Maximum Likelihood decoding is a central yet computationally challenging problem in coding theory. It was proven to be \textsf{NP}-complete for general linear codes by Berlekamp \textit{et al.} in 1978~\cite{NP_hardness_general_linear,Minimum_distance_hardness}, and this hardness extends even to highly structured code families like Reed--Solomon codes~\cite{RS_NP_hardness}. The straightforward brute-force solution, known as exhaustive search decoding (ESD), provides a simple complexity baseline for ML decoding.

In a typical setting of binary coding over a discrete memoryless channel (DMC), the sender transmits a codeword 
from a binary code \(\mathscr{C}[n, k]_2\) with $2^k$ codewords, and the receiver observes a possibly corrupted sequence \(\boldsymbol{y} \in \mathbb{F}_2^n\). ESD identifies the ML codeword by computing the likelihood \(P(\boldsymbol{y} \mid \boldsymbol{c})\) for all \(\boldsymbol{c} \in \mathscr{C}\) and selecting the maximizer. This procedure requires \(n\) operations per codeword, leading to a \emph{time complexity} of \(\mathcal{O}(2^kn)\). Although parallel hardware can evaluate many likelihoods simultaneously and thereby reduce wall-clock time, the overall number of multiplications---and hence the fundamental time complexity--remains unchanged. For this reason, complexity comparisons are conventionally made in terms of time complexity, independent of parallelization. Alongside time complexity, one also considers \emph{space complexity}, which measures the amount of memory required by an algorithm as a function of the input size. In the case of ESD, the algorithm evaluates one codeword at a time and therefore needs only $\mathcal{O}(n)$ memory to store this codeword, making it space-efficient despite its exponential time complexity.

When the noise model is known, the likelihood measure of memoryless channels equivalently reduces to a distance metric—for instance, Hamming distance for memoryless symmetric channels with bit-flip noise, and Euclidean distance for additive white Gaussian noise (AWGN) channels. For example, in a binary symmetric channel (BSC), ML decoding reduces to minimum-distance decoding: the most likely codeword is the one that is closest in Hamming distance to the received sequence. For linear codes, a refinement is possible. When the code rate is $R=k/n \leq 1/2$, ESD involves checking all $2^k$ codewords. When $R>1/2$, it is more efficient to consider the $2^{n-k}$ possible \emph{syndromes} and perform \emph{syndrome decoding}, which is also a minimum distance decoder and thus ML. In summary, the complexity of ESD for binary linear codes is well known to be $\mathcal{O}(2^{\min\{k,\, n-k\}}n)$~\cite{ESD_decoding}. We emphasize that this refinement applies only to linear codes, since syndrome decoding is not defined for nonlinear codes. A key characteristic of the ESD is that it is \emph{universal}\footnote{In Information theory, the term \emph{universal decoding} typically refers to decoding without explicit knowledge of the channel law~\cite{GRAND_random,miyamoto2025universal}. In this work, however, we use \emph{universal} to mean that the decoder is not tied to any specific codebook, consistent with its usage in, for example,~\cite{Universal_decoder}. In any case, the intended meaning should be clear from the context~\cite{duffy2025personal}.} (i.e., \emph{code-agnostic}) as it does not exploit any algebraic structure of the code, and thus its performance depends only on the code size rather than its specific construction.

\subsection{Brief Historical Prospective}
Due to the prohibitive complexity of ESD (linear in block length $n$ and exponential in code dimension $k$), significant effort has been invested in developing more efficient ML or near-ML decoders. We provide a broad classification into a few main categories listed in Table~\ref{tab:decoder_comparison_concise}.

An early approach, introduced by Wolf in his seminal 1978 paper~\cite{ViterbiForBlock:Wolf'78}, demonstrated that ML decoding of any \emph{linear} block code can be carried out using the Viterbi algorithm~\cite{Viterbi}—(originally devised for convolutional codes) on a trellis with at most $2^{n-k}$ states and $2^k$ distinct paths of depth $n$. This trellis-based algorithm yields a time complexity of $\Oh(2^{\min\{k,\, n-k\}}n)$, which essentially matches that of ESD, and a space complexity of at most $\Oh(2^{n-k}n)$. The algorithm's importance lies in placing ML decoding for general linear codes under a trellis framework, enabling efficient and hardware-friendly implementations, and revealing that some codes (e.g., convolutional codes) admit compact trellis representations where Viterbi decoding is significantly more efficient. The Wolf's perspective led to the recognition of \emph{trellis complexity} as a fundamental code parameter, alongside length, dimension, and minimum distance~\cite{trellis_complexity_binary,trellis_decoding_complexity}. Extending this framework, Forney showed that the Viterbi algorithm can also be applied to ML decoding of linear block codes transmitted over $L$-tap ISI channels~\cite{ML_sequence_estimator_Viterbi}, where $L$ denotes the channel memory, i.e., the number of past symbols that affect each received symbol. In this setting, the decoding complexity scales proportionally to $2^L$. Notably, this remains the only known approach to exact ML decoding for ISI channels. The well-known BCJR algorithm~\cite{BCJR_algorithm} performs symbol-optimal detection, minimizing the symbol error rate rather than the word error rate; thus, it is not a sequence ML decoding algorithm; in fact, the optimally decoded output need not be a valid codeword.

Another approach reformulates the ML decoding problem over a discrete memoryless channel
symmetric channel (-hard in general) as an integer linear program (ILP), whose optimal solution corresponds to the ML codeword; see~\cite{ML_decoding_Linear_program} and related subsequent works. Relaxing the integrality constraints yields a linear program (LP) that can be solved in \emph{polynomial time}, for instance by the Ellipsoid algorithm~\cite{Linear_optimization_textbook}. Under certain conditions, the LP relaxation is \emph{tight}, meaning its solution coincides with that of the original ILP and hence produces the ML codeword in polynomial time. When these conditions fail, however, the LP solution is typically fractional and does not correspond to any valid codeword. In short, this method does not guarantee recovery of the ML codeword in general. 

Practical coding standards use particular decoders co-designed with code-specific decoding methods to achieve an ML decoding performance in polynomial time for a restricted class of codes or channels. Prominent examples include Bose–Chaudhuri–Hocquenghem (BCH) codes decoded by the Berlekamp–Massey algorithm~\cite{Berlekamp_Massey_algorithm}, decoding of expander codes over the BSC~\cite{expander_code_bsc}, and decoding of first-order Reed-Muller codes~\cite{MLD_RM_1st}. Recently, code-agnostic approaches to ML decoding have attracted significant attention, as they enable universal applicability across arbitrary codes and support practical implementations in reliable communication settings, where the SNR is typically high and moderate-to-high-rate codes are employed. This line of research was initiated by the remarkable work of Duffy et al.~\cite{GRAND:DuffyKLM'19}, who introduced the method of ML decoding via \emph{noise guessing}, termed GRAND. This framework addresses discrete memoryless additive-noise channels, where the received sequence equals the transmitted codeword plus noise (modulo the alphabet). Because the noise uniquely determines the transmitted codeword given the received sequence, decoding reduces to testing (or, \emph{guessing}) candidate noise vectors ranked by likelihood. Each is subtracted from the received sequence and checked for codebook membership; the first valid residual yields the ML codeword. The computational cost is governed by the number of guesses required before termination. Although an exact complexity is unknown, it was shown that in the asymptotic regime $n \to \, \infty$, the expected number of guesses scales as
$2^{\,n \min \{ H_{1/2},\, 1 - \frac{k}{n} \}}$,
where $H_{\alpha}$ denotes the R\'enyi entropy of order $\alpha$ of the noise (with $H_1$ being the Shannon entropy). As a Viterbi decoding approach for linear block codes, this method offers substantially lower complexity than ESD in the high-SNR regime (i.e., when noise entropy is small) and is particularly effective for high-rate, short block codes. Its advantage, however, vanishes as the SNR decreases, the code rate is reduced, and especially when the channel is non-additive.

Several techniques have been proposed to reduce the complexity of ML decoding in the high-SNR regime, typically at the cost of a slight increase in error probability. Such methods are therefore commonly referred to as \emph{approximate} ML decoding, or \emph{near-optimum decoding}. While asymptotically optimal at high SNR, these methods perform worse than exact ML decoding at low SNR or for practical block lengths, and the gap in error performance compared to MLD increases with the code dimension. Nevertheless, the worst-case complexity (e.g., in low-SNR conditions) remains essentially the same as that of ESD; see, for example,~\cite{Chase_algo,BinaryJ_ary:Dorsch74,order_statistic_decoding,wonterghem2017OSD}. 

As a related note, extensions of ML decoding to list decoding, soft-decision decoding (or simply \emph{soft decoding}), and erasure decoding have been widely studied. For example, Elias’s early work on list decoding (also called as list-$\ell$ decoding, where $\ell$ denotes the list size)~\cite{elias1957list} showed that allowing the decoder to output multiple candidates (i.e., allow some level of \emph{ambiguity}) can reduce error probability, while erasure decoding provides a simple way to incorporate partial reliability information. Soft-decision decoding, particularly with ML-based algorithms, has since become the standard in practice~\cite{Generalized_distance_decoding}. However, each of these approaches is tailored to a specific setting, and no existing method provides a unified, code-agnostic framework that applies simultaneously to ML hard-decision, soft-decision, erasure, and list decoding.


\subsection{The Proposed Algorithm and its Relation to the Prior Work}


This paper introduces a code-agnostic framework that, for the first time, breaks the long-standing \(\mathcal{O}(2^kn)\) complexity barrier of ESD. The key idea is to express the likelihood of each codeword as the inner product of two vectors: one derived from the received sequence, and the other representing the codeword itself. This allows us to represent the entire codebook as a single, large, binary matrix \(\mathbf{M}\). Consequently, ML decoding is reduced to a single vector–matrix multiplication, followed by selecting the maximum entry of the resulting vector.

The efficiency of the proposed approach stems from exploiting the discrete structure of $\mathbf{M}$. This performance gain parallels the classical distinction between sorting arbitrary real numbers and sorting integers. In general, sorting $n$ arbitrary items requires $\Omega(n \log n)$ comparisons (i.e., grows on the order of \(n\log n\)), whereas algorithms such as Counting Sort—particularly efficient when the range of possible input values is small relative to the number of elements—leverage the finite alphabet of integers to achieve linear $\mathcal{O}(n + k)$ complexity~\cite{Knuth1998}, where $k$ denotes the range size of the integer-valued inputs. Likewise, rather than treating the entries of \(\mathbf{M}\) as arbitrary reals, we exploit its binary structure via the Mailman algorithm~\cite{mailman_algorithm}, which computes the product of an \(m \times n\) binary matrix \(\mathbf{A}\) with a real vector \(\boldsymbol{x} \in \mathbb{R}^n\) in \(\mathcal{O}(mn / \log \max\{m, n\})\) amortized time complexity—improving upon the straightforward \(\mathcal{O}(mn)\) complexity for general matrices. This contrasts with the classical result of Winograd~\cite{winograd1968number}, which establishes that general vector-matrix multiplication requires \(\Omega(mn)\) operations. Employing this algorithm reduces the total number of operations to \(\mathcal{O}(2^k)\)—an \(n\)-fold improvement over ESD. For the special case of linear codes over the BSC, the framework further adapts to syndrome decoding with a complexity of \(\mathcal{O}(2^{\min\{k,\, n-k\}})\), again improving upon the standard \(\mathcal{O}(n \cdot 2^{\min\{k,\, n-k\}})\) complexity of trellis-based ML decoders.
Since the time complexity is inherently proportional to the number of codewords (or syndromes, whichever is smaller), no further substantial reduction should be expected—identifying the most likely codeword necessarily entails examining each candidate at least once. The main drawback lies in memory usage: the method stores a matrix representing all codewords in the codebook, resulting in a space complexity of \( \mathcal{O}(2^{k}) \). Nonetheless, this is still smaller by a factor of \( n \) compared to trellis-based (Viterbi) decoding. Moreover, the framework is universal: it applies to both linear and nonlinear block codes, supports hard- and soft-decision decoding, and extends naturally to general memoryless as well as intersymbol-interference (ISI) channels—the most widely studied class of channels with memory~\cite{pfister2003phd}.

The idea of representing codeword likelihoods as inner products can be traced back, in an early form, to the work of Conway and Sloane in 1986~\cite{DecodingTechnique:ConwayS'86}, who studied specific channels with binary codes and showed that, in those cases, ML decoding is equivalent to finding the codeword closest in Euclidean distance. By contrast, our method makes no assumptions about the noise model and is not a distance-based decoder; it can thus be viewed as a systematic and efficiently implementable generalization of their approach.





Unlike all previously known ML decoding methods, the proposed decoder universally achieves a complexity of $\Oh(2^k)$, which is independent of the block length $n$ when the code rate $k/n$ is assumed to be fixed. This complexity is also independent of the SNR level, distinguishing it from approaches like GRAND~\cite{GRAND:DuffyKLM'19}. Moreover, the algorithm does not depend on the specific received sequence, unlike the algebraic ML decoder of~\cite{MLAlgebraicDecoder:KanekoNIH'94}. The complexity of our proposed method is smaller by a logarithmic factor compared to the best known trellis-based ML decoding via the Viterbi algorithm. This makes the method particularly attractive for low-SNR channels—a relatively unexplored regime where reliable transmission requires low-rate codes with large minimum distance~\cite{low_capacity_RM}. For example, Reed–Muller (RM) codes \( \text{RM}(r, m) \) have long been employed in deep-space and satellite communications due to their high minimum distance and simple majority-logic decoding~\cite{Marriage_heaven,ReedMuller_benchmark,One_step_RM}. For fixed order \( r \), the dimension of RM codes scales as \( k = \Omega(\log n) \), i.e., it grows logarithmically with the block length \( n \). In this regime of very low SNRs and code rates, approximate ML decoding algorithms fail to provide satisfactory performance, and exact ML decoders cannot achieve lower complexity than exhaustive search. Moreover, when extended to list decoding, the proposed algorithm’s complexity increases only logarithmically with the list size, unlike other approaches such as SOGRAND~\cite{SOGRAND:YuanMGD'25} and List Viterbi~\cite{ListViterbi:SeshadriS'94}, whose complexity grows linearly with the list size.

Our approach (just as ESD and Viterbi decoding) does not impose assumptions on the noise or channel model. In contrast, prior works such as~\cite{GRAND:DuffyKLM'19,UniVersalList:XiangpingX'25} assume additive noise, with~\cite{GRAND:DuffyKLM'19} focusing on symmetric channels and~\cite{UniVersalList:XiangpingX'25} restricted to discrete (binary-input, multiple-output) channels. Our method achieves these advantages at the expense of increased storage, i.e., higher \emph{space complexity}. This observation is consistent with the well-known principle of \emph{time--space tradeoffs} in computer science, where reductions in computational time often require increased memory usage~\cite{CormenEtAl2009}.

The proposed ML decoding framework is particularly well suited for ultra-reliable low-latency communication (URLLC), a foundational requirement of 5G and emerging 6G systems. In conventional universal decoders, improving reliability by increasing the block length \(n\) typically incurs a proportional increase in decoding latency, often scaling as \(\mathcal{O}(2^k n)\). In contrast, the proposed method achieves an \(n\)-fold reduction in complexity, yielding \(\mathcal{O}(2^k)\) decoding time and effectively decoupling latency from the block length. This allows system designers to increase redundancy and reliability without incurring latency penalties, offering a clear advantage over iterative or trellis-based decoders. Owing to its deterministic runtime and reliance on structured vector--matrix multiplication, the framework is well suited for efficient implementation on parallel hardware such as GPUs or FPGAs~\cite{harris2013digital}. These features position the proposed decoder as a promising enabler for latency-critical applications that demand both high reliability and ultra-low latency, including the Tactile Internet, remote surgery, AR/VR/XR, vehicle-to-everything (V2X) communication, and industrial automation~\cite{Le_URLLC_Physics,Maghsoudnia2024,Yue2023URLLC}.




\begin{table*}[t]
\centering
\caption{Comparison of Exact ML Decoding Approaches for General Linear $[n, k]_2$ Codes.}
\label{tab:decoder_comparison_concise}
\begin{tabular}{@{}llll@{}}
\toprule
\textbf{Decoding Method} & \textbf{Time Complexity} & \textbf{Space Complexity} & \textbf{Characteristics} \\ \midrule
Exhaustive Search (ESD) & $\mathcal{O}(2^kn)$ in general & $\Oh(n)$ & Code-agnostic; Works for all codes for both hard- and\\
& $\Oh(2^{\min\{k,\, n-k\}}n)$ for linear codes & & soft-decision decoding. Prohibitive time complexity. \\ \hline
Trellis-based (Viterbi)~\cite{ViterbiForBlock:Wolf'78} & $\mathcal{O}(2^{\min\{k,\, n-k\}}n)$ for linear codes & $\Oh(2^{n-k}n)$ & Code-agnostic; Works for linear codes only.\\ 
 & & & Same time complexity as ESD. High space complexity.\\
\hline
LP Relaxation~\cite{ML_decoding_Linear_program} & Polynomial in $n$ & & Fails if relaxation is not tight. \\ 
& & & Works for discrete memoryless symmetric channels only. \\
\hline
GRAND~\cite{GRAND:DuffyKLM'19} & $\Oh(2^{\,n \min \{ H_{{1}/{2}},\, 1 - \frac{k}{n} \}}n)$ & $\Oh(n)$ & Low complexity under high SNRs. Otherwise, same as ESD.\\ 
 & (Asymptotically) & & Requires additive-noise model. \\
\hline
Algebraic Decoders~\cite{Berlekamp_Massey_algorithm,expander_code_bsc,MLD_RM_1st} & Polynomial in $n$ & & Applicable to particular codes/channels only. \\ \midrule
\emph{Our method}. In general & $\mathcal{O}(2^k)$ & $\Oh(2^{k})$& Code-agnostic; Works for all codes for both hard- and\\
\hspace{0.57in} For linear codes & $\Oh(2^{\min\{k,\, n-k\}})$ & $\Oh(2^{\min\{k,\, n-k\}})$ & soft-decision decoding. High space complexity. \\
\bottomrule
\end{tabular}
\end{table*}

\subsection{Organization}
The core contributions of this paper are presented in Section~\ref{sec:formulation}. Part~\ref{subsec:discrete} introduces the universal decoding method for discrete memoryless channels and analyzes its complexity. Part~\ref{subsec:soft_list} extends the framework to soft-decision decoding and list-$\ell$ ML decoding, and derives the associated complexities. Part~\ref{subsec:erasure_channel} applies the method to erasure decoding. Part~\ref{subsec:linear_syndrome} focuses on linear codes over symmetric channels, showing how syndrome decoding can be used to further reduce complexity. Part~\ref{subsec:ISI} then treats the application of the proposed framework to channels with memory, namely, $L$-tap ISI channels. Section~\ref{sec:Simulation} provides simulation results validating
the theoretical analysis. We conclude the paper in Section~\ref{sec:Conclusion}, where we also discuss the inherent difficulty of the underlying vector-matrix multiplication routine and, drawing on prior work—though without formal proof—argue that further substantial improvements in complexity are unlikely. Finally, a brief overview of the Mailman algorithm is given in Appendix~\ref{appendix:mailman}. 

\section{Formulation}\label{sec:formulation}
In the rest of this paper, the finite field over a prime or prime power \(q\) is denoted as \(\mathbb{F}_q\). A \(q\)-ary \textbf{linear code} \(\mathscr{C}\) with parameters \([n, k, d]_q\) is a \(k\)-dimensional subspace of the vector space \(\mathbb{F}_q^n\) with minimum distance \(d\). More generally, any set of \(S\) codewords in \(\mathbb{F}_q^n\) with minimum distance \(d\) forms a code, which may be non-linear. The logarithm, denoted by \(\log\), is taken to base~2 unless stated otherwise.


\subsection{Decoding Algorithm Formulation}\label{subsec:discrete}
We denote a generic discrete memoryless channel (DMC) by \( P: \mathcal{X} \rightarrow \mathcal{Y} \), where \( \mathcal{X} \) and \( \mathcal{Y} \) are the input and output alphabets, respectively. For example, in the BSC, $\mathcal{X} = \mathcal{Y} = \{0, \, 1\}$, while for the AWGN channel, $\mathcal{Y} = \mathbb{R}$. The channel is characterized by the conditional transition probabilities \( P_{Y|X}(y \mid x) \) for each \( x \in \mathcal{X} \) and \( y \in \mathcal{Y} \). While the output alphabet \( \mathcal{Y} \) may be arbitrary, we assume the input alphabet \( \mathcal{X} \) is discrete. This is because in digital communication, information is first encoded using a channel code defined over a finite alphabet before modulation and transmission. Without loss of generality, we index the input symbols as \( \mathcal{X} = \{1, 2, \dots, q\} \), where \( q = |\mathcal{X}| \) denotes the alphabet size. A $q$-ary block code (not necessarily linear) $\mathscr{C}[n, k]_q \subseteq \mathcal{X}^n$ of block length $n$ and information length $k$ (i.e., the number of information symbols) is assumed to be shared between the sender and the receiver. When the code is linear, $k$ is also referred to as the \emph{dimension} of the code. Suppose a codeword \( \boldsymbol{x} \in \mathscr{C} \) is transmitted over the channel, resulting in a received sequence \( \boldsymbol{y} \in \mathcal{Y}^n \). Throughout this paper, we index vector components from \( 1 \) to their length, and use \( y_i \) and \( \boldsymbol{y}[i] \) interchangeably to denote the \( i \)-th component of \( \boldsymbol{y} \). For a vector $\boldsymbol{v}$, its transpose is denoted by $\boldsymbol{v}^{\top}$.

We write \( P_{Y|X}^n \) to denote the product channel corresponding to \( n \) independent uses of the memoryless channel \( P \), i.e.,
\[
P_{Y|X}^n : \mathcal{X}^n \rightarrow \mathcal{Y}^n, \quad P_{Y|X}^n(\boldsymbol{y} \mid \boldsymbol{x}) = \prod_{i=1}^n P_{Y|X}(y_i \mid x_i).
\]

For simplicity, we often write \( P \) instead of \( P_{Y|X} \) when the context is clear. ML decoding over a DMC seeks the codeword \( \boldsymbol{c}^* \in \mathscr{C} \) that maximizes the conditional probability of the received sequence \( \boldsymbol{y} \in \mathcal{Y}^n \), namely
\begin{align}
    \boldsymbol{c}^* 
    &= \arg \max_{\boldsymbol{c} \in \mathscr{C}} P^n_{Y|X}(\boldsymbol{y} \mid \boldsymbol{c})
    = \arg \max_{\boldsymbol{c} \in \mathscr{C}} \log \left( \prod_{i=1}^n P(y_i \mid c_i) \right) \notag\\
    &= \arg \max_{\boldsymbol{c} \in \mathscr{C}} \sum_{i=1}^n \log P(y_i \mid c_i).
    \label{eq:ML_decoding_dmc}
\end{align}

ML decoding guarantees the minimum probability of decoding error when the \textit{a priori} probabilities of all the codewords are equal~\cite{Proakis2008Digital}. To facilitate this computation, we define the \emph{(log) conditional probability vector} 
\( V(\boldsymbol{y}) \in \mathbb{R}^{qn} \), which records the logarithmic 
transition probabilities across all possible input symbols 
\( x_i \in \mathcal{X} = \{1, \dots, q\} \). Specifically,
\begin{align}
V(\boldsymbol{y}) = \left[ \boldsymbol{v}_1(\boldsymbol{y}),\, \boldsymbol{v}_2(\boldsymbol{y}),\, \dots,\, \boldsymbol{v}_n(\boldsymbol{y}) \right],
\label{eq:log_likelihood_vector_dmc}
\end{align}
where each sub-vector \( \boldsymbol{v}_i(\boldsymbol{y}) \in \mathbb{R}^q \) stores the log-likelihoods
of receiving \( y_i \) given every possible input symbol
\begin{align} 
\boldsymbol{v}_i(\boldsymbol{y}) = \big[ &\log P(y_i \mid c_i = 1),\, \log P(y_i \mid c_i = 2),\, \dots \notag\\ &\log P(y_i \mid c_i = q-1),\, \log P(y_i \mid c_i = q) \big]. \label{eq:log_likelihood_vector_dmc_component} 
\end{align}



We also define the \emph{incidence vector} \( I(\boldsymbol{c}) \in \{0, 1\}^{qn} \),
constructed from each codeword \( \boldsymbol{c} \in \mathcal{X}^n \), as
\begin{align}
I(\boldsymbol{c}) = \left[ \boldsymbol{u}_1(\boldsymbol{c}),\, \boldsymbol{u}_2(\boldsymbol{c}),\, \dots,\, \boldsymbol{u}_n(\boldsymbol{c}) \right],
\label{eq:incidence_vector_dmc}
\end{align}
where each sub-vector \( \boldsymbol{u}_i(\boldsymbol{c}) \in \{0, 1\}^q \) is the one-hot encoding of \( c_i \), that is
\[
\boldsymbol{u}_i(\boldsymbol{c}) = \big[ \mathds{1}\{c_i = 1\},\, \mathds{1}\{c_i = 2\},\, \dots,\, \mathds{1}\{c_i = q\} \big],
\]
with \( \mathds{1}\{\cdot\} \) denoting the indicator function (\( \mathds{1}\{A\} = 1 \) if event \( A \) is true, and 0 otherwise).  
Each \( \boldsymbol{u}_i(\boldsymbol{c}) \) therefore has a single 1 in the coordinate matching the symbol \( c_i \), and 0s elsewhere. Consequently, \(I(\boldsymbol{c})\) is a binary vector of length \(nq\) with exactly \(n\) ones. With this notation, MLD rule in Eq.~\eqref{eq:ML_decoding_dmc} presents as 
\[
\boldsymbol{c}^* = \arg \max_{\boldsymbol{c} \, \in \, \mathscr{C}} \sum_{i=1}^n\boldsymbol{v}_i(\boldsymbol{y}) \cdot \boldsymbol{u}_i(\boldsymbol{c})^\top = \arg \max_{\boldsymbol{c} \, \in \, \mathscr{C}} V(\boldsymbol{y}) \cdot I(\boldsymbol{c})^\top,
\]
meaning that the likelihood of each codeword is given by the inner product of two vectors. The first, \( V(\boldsymbol{y}) \), is a conditional probability vector determined solely by the received sequence \( \boldsymbol{y} \) and the channel transition probabilities \( P(\boldsymbol{y} \mid \boldsymbol{x}) \), and is independent of the codebook (it only depends on the alphabet of the codebook). The second, \( I(\boldsymbol{c}) \), is a codeword-specific incidence vector that is independent of both the received sequence and the channel model.

Let \( S = |\mathscr{C}| \) denote the \emph{size of the codebook}, i.e., the number of codewords, and $[S] = \{1, 2, \dots, S\}$. We denote the codewords be indexed as \( \boldsymbol{c}^1, \boldsymbol{c}^2, \dots, \boldsymbol{c}^S \). Since each codeword $\boldsymbol{c}^j$ is uniquely associated with a codeword index $j$, ML decoding then amounts to finding
\[
i^* = \arg \max_{i \, \in \, [S]} V(\boldsymbol{y}) \cdot I(\boldsymbol{c}^i)^\top.
\]
We collect the incidence vectors of all codewords as columns of a binary matrix
\begin{align}
    \mathbf{M} = \big[ I(\boldsymbol{c}^1)^\top \;\; I(\boldsymbol{c}^2)^\top \;\; \cdots \;\; I(\boldsymbol{c}^S)^\top \big] \, \in \, \{0, 1\}^{nq^{} \times S}.
    \label{eq:codebook_matrix_isi}
\end{align}
The resulting log-likelihood vector is given by
\[
\mathbf{m} = V(\boldsymbol{y}) \cdot \mathbf{M} \ \ \in \ \ \mathbb{R}^S, 
\]
where 
\begin{align}
\mathbf{m}[i] = \log P(\boldsymbol{y} \mid \boldsymbol{c}^i),\, \forall\, i \, \in \, [S],
\label{eq:log_likelihood_vector}
\end{align} 
is the log likelihood of the codeword $\boldsymbol{c}^i$. Thus, ML decoding reduces to identifying the index of the maximum entry in a length-\( S \), real-valued vector
\[
i^* = \arg \max_{i \, \in \, [S]} \mathbf{m}[i],
\]
which can be accomplished using a trivial linear-time scan: by iterating through the vector once and updating a register that stores the current maximum value. Since the length of the resulting vector $\mathbf{m}$ is $S$, this step has a computational complexity of \( \Oh(S) \). We summarize the procedure in Box~\ref{box:decoding}, then analyze the overall decoding complexity.
\begin{myrefbox}[box:decoding]{ML decoding via Vector–Matrix Multiplication}
\textbf{Pre-processing:} For a code \( \mathscr{C} = \{\boldsymbol{c}^1, \boldsymbol{c}^2, \dots, \boldsymbol{c}^S\} \), construct the binary matrix \( \mathbf{M} \) as defined in Eq.~\eqref{eq:codebook_matrix_isi}.
\begin{itemize}
    \item \textbf{Step 1: Construct the conditional probability vector.}  
    Given a received sequence \( \boldsymbol{y} \), build the conditional probability vector \( V(\boldsymbol{y}) \) according to Eq.~\eqref{eq:log_likelihood_vector_dmc}  and~\eqref{eq:log_likelihood_vector_dmc_component}.
    
    \item \textbf{Step 2: Perform vector–matrix multiplication.}  
    Compute the score vector \( \mathbf{m} = V(\boldsymbol{y}) \cdot \mathbf{M} \, \in \, \mathbb{R}^S \).
    
    \item \textbf{Step 3: Identify the maximum-likelihood codeword.}  
    Let \( i^* = \arg\max\limits_{i \, \in \, [S]} \mathbf{m}[i] \). Then the ML codeword is
    \[
    \boldsymbol{c}^* = \boldsymbol{c}^{i^*}.
    \]
\end{itemize}
\end{myrefbox}
\emph{Complexity Analysis.} We now analyze the time complexity and space complexity of the algorithm.
    
\textbf{Time complexity}. The overall complexity is obtained by summing the cost of the three following steps

\begin{itemize}
    \item \textbf{Step 1} requires constructing the conditional probability vector \( V(\boldsymbol{y}) \in \mathbb{R}^{qn} \), which has complexity \( \Oh(nq^{}) \).
    
    \item \textbf{Step 2} involves a vector–matrix multiplication between a real-valued vector $V(\boldsymbol{y})$ of length \( nq \) and a binary matrix $\mathbf{M}$ of size \( nq \times S \). The matrix $\mathbf{M}$ by our construction has a specific structure: it is not only binary but also highly sparse, containing exactly $n$ ones per column. Its nature allows this product to be computed efficiently using the Mailman algorithm~\cite{mailman_algorithm} (see Appendix~\ref{appendix:mailman} for details), with complexity
    \begin{align}\label{eq:mailman_complexity}
    \Oh\left(\dfrac{nq \cdot S}{\log\left(\max\{nq, S\}\right)}\right). 
    \end{align}
    For a \( q \)-ary linear code \( \mathscr{C} \) of dimension \( k \) with \( S = q^k \) codewords, and assuming \( \max\{nq, q^k\} = q^k \), the complexity simplifies to
    \[
    \Oh\left(\dfrac{nq \cdot q^k}{k \log q}\right) = \Oh\left(\dfrac{1}{R} \cdot \dfrac{q}{\log q} \cdot q^k\right) = \Oh(q^k),
    \]
    where \( R = k/n \) denotes fixed code rate of the particular code under consideration. The term \(q/\log q\) is also treated as a constant for any practical code, e.g., for binary codes, \(q/\log_2 q = 2\).
    
    \item \textbf{Step 3} finds the maximum entry in the score vector \( \mathbf{m} \) of length $S$ via a linear-time scan, with complexity \( \Oh(S) = \Oh(q^k) \) for a \( q \)-ary linear $[n, k]_q$ code. Note that since the time complexity is already proportional to the number of codewords $S$, no further substantial reduction is expected, as identifying the most likely codeword requires examining each candidate at least once.
\end{itemize}
Hence, the total decoding time complexity is 
\[
\Oh(nq^{}) + \Oh(q^{k}) + \Oh(q^k) = \Oh(q^{k}),
\]
since, for practical block codes, the number of information symbols $k$ increases roughly in proportion to the block length $n$, making the term $nq$ negligible compared to $q^{k}$.

\textbf{Space complexity.} Step 2 of our algorithm employs the Mailman algorithm~\cite{mailman_algorithm} for fast vector–matrix multiplication by factoring the codebook matrix as \( \mathbf{M} = UP \) (see Appendix~\ref{appendix:mailman}). As shown therein, only the permutation matrix \( P \), which is a binary matrix of size $S \times S$, needs to be stored at the receiver, and its particular structure allows efficient representation with space complexity \(\Oh(S)= \mathcal{O}(q^k) \).

\begin{remark}
The matrix \( \mathbf{M} \) and its factorization \( \mathbf{M} = UP \) are constructed once at the receiver as an offline pre-processing step. Hence, their computational costs are amortized and excluded from the overall complexity analysis. This treatment follows the convention in prior works, e.g.,~\cite{UniVersalList:XiangpingX'25}, and parallels the one-time trellis construction required by the Viterbi algorithm~\cite{ViterbiForBlock:Wolf'78}.
\end{remark}

\subsection{Extension to Soft Decoding and List Decoding}
\label{subsec:soft_list}
We now demonstrate that the proposed method extends naturally to both (maximum likelihood) soft decoding and list decoding.

\subsubsection*{Soft Decoding}  
Observe that the output alphabet $\mathcal{Y}$ and the (conditional) transition probabilities can be arbitrary. Particularly, if the received vector \( \boldsymbol{y} \) is considered \emph{prior to demodulation}—for instance, as the real-valued output of a matched filter—then the output alphabet becomes continuous, i.e., \( \mathcal{Y} = \mathbb{R} \). This scenario corresponds to \emph{soft decoding}, where the channel transition probability $P_{Y|X}$ becomes a conditional probability \emph{density function} $p_{Y|X}$. It is well known—see, for example,~\cite[Ch. 4]{Proakis2008Digital}—that ML decoding in this setting is
\begin{align*}
\boldsymbol{c}^* 
& = \arg \max_{\boldsymbol{c} \, \in \, \mathscr{C}} p_{Y|X}(\boldsymbol{y} \mid \boldsymbol{c}) = \arg \max_{\boldsymbol{c} \in \mathscr{C}} \log \left( \prod_{i=1}^n p_{Y|X}(y_i \mid c_i) \right) \notag\\
    &= \arg \max_{\boldsymbol{c} \in \mathscr{C}} \sum_{i=1}^n \log p_{Y|X}(y_i \mid c_i).
\end{align*}
that is, the task of finding the codeword \( \boldsymbol{c} \) that maximizes the likelihood \( p_{Y|X}(\boldsymbol{y} \mid \boldsymbol{c}) \). This is structurally equivalent to the standard ML decoding formulation given in Eq.~\eqref{eq:ML_decoding_dmc} and can therefore be handled using the same approach.

\subsubsection*{List Decoding} In unique ML decoding, the goal is to find the codeword \( \boldsymbol{c}^* \, \in \, \mathscr{C} \) that maximizes \( P_{Y|X}(\boldsymbol{y} \mid \boldsymbol{c}) \). In list-\( \ell \) ML decoding, where the list size $\ell$ is fixed, the objective is to return a list of \( \ell \ge 1 \) codewords whose likelihoods are the \( \ell \) largest among all codewords in the codebook. Formally, one seeks a list \( \mathcal{L}_\ell \subseteq [S] \) such that
\[
\mathcal{L}_\ell = \{ i_1, \dots, i_\ell \} \text{ with } 
\mathbf{m}[i_j] \ge \mathbf{m}[h],\quad \forall\, h \, \in \, [S]\setminus\mathcal{L}_\ell,
\]
\[
\text{and } \mathbf{m}[i_1] \ge \mathbf{m}[i_2] \ge \dots \ge \mathbf{m}[i_\ell].
\]
This reduces to the problem of selecting and sorting the \( \ell \) largest elements from an unordered list of \( S \) values. This task can be done with complexity \( \Oh(S \log \ell)\) by using, for instance, a Min-heap–based selection algorithm~\cite{GeeksforGeeksKLargest}. The time complexity grows only logarithmically with the list size $\ell$.
\subsection{Application to Erasure Channels}
\label{subsec:erasure_channel}

The proposed framework extends naturally to binary erasure channels. On such a channel, the goal is to find the unique codeword \(\boldsymbol{c}^*\) that is consistent with the received sequence \(\boldsymbol{y}\) on all unerased positions, a task that is guaranteed to have a unique solution if the number of erasures \(n_\epsilon\) is less than the code's minimum distance \(d\).

The key adaptation is to use a bipolar encoding scheme instead of the one-hot encoding used for the DMC. Let \(\mathcal{K}\) be the set of known (i.e., unerased) positions. We define a vector based on the received sequence \(\boldsymbol{y} \in \{0,1,\epsilon\}^n\) and a vector for each codeword \(\boldsymbol{c} \in \{0,1\}^n\) as follows:
\[
V(\boldsymbol{y})[i] = 
\begin{cases}
    1, & \text{if } y_i = 1 \\
   -1, & \text{if } y_i = 0 \\
    0, & \text{if } y_i = \epsilon
\end{cases}
~\text{and}~
I(\boldsymbol{c})[i] = 
\begin{cases}
    1, & \text{if } c_i = 1 \\
   -1, & \text{if } c_i = 0
\end{cases}
\]
With this representation, the product $V(\boldsymbol{y})[i] \, I(\boldsymbol{c})[i]$ equals $0$ only when $y_i$ is erased; otherwise, it equals $1$ if $y_i$ and $c_i$ agree, and $-1$ if they differ. The inner product then has a remarkable property. It directly computes a score that is maximized by the codeword with the fewest disagreements on the unerased positions with sequence $\boldsymbol{y}$. Let \(d_{\mathcal{K}}(\boldsymbol{y}, \boldsymbol{c})\) be the number of positions in \(\mathcal{K}\) where \(y_i \neq c_i\), then $|\mathcal{K}| - d_{\mathcal{K}}(\boldsymbol{y}, \boldsymbol{c})$ is the number of positions that $\boldsymbol{y}$ and $\boldsymbol{c}$ match. The inner product evaluates to
\begin{align}
    V(\boldsymbol{y}) \cdot & I(\boldsymbol{c})^\top =\nonumber\\
    & (|\mathcal{K}| - d_{\mathcal{K}}(\boldsymbol{y}, \boldsymbol{c})) - d_{\mathcal{K}}(\boldsymbol{y}, \boldsymbol{c}) = |\mathcal{K}| - 2d_{\mathcal{K}}(\boldsymbol{y}, \boldsymbol{c}).
    \label{eq:erasure_inner_product}
\end{align}
Maximizing this score is equivalent to minimizing \(d_{\mathcal{K}}(\boldsymbol{y}, \boldsymbol{c})\). The unique correct codeword \(\boldsymbol{c}^*\) has \(d_{\mathcal{K}}(\boldsymbol{y}, \boldsymbol{c}^*) = 0\), which yields the maximum possible score of \(|\mathcal{K}| = n - n_\epsilon\).

The decoding procedure is therefore identical to the main formulation. For a code $\mathscr{C}$ of $S$ codewords, we first pre-compute the codebook matrix \(\mathbf{M} = [I(\boldsymbol{c}^1)^\top, \dots, I(\boldsymbol{c}^S)^\top]\) over \(\{-1, 1\}^{n \times S}\), and for a given \(\boldsymbol{y}\), we compute the score vector:
\[
\mathbf{m} = V(\boldsymbol{y}) \cdot \mathbf{M}.
\]
The index of the decoded codeword is simply \(i^* = \arg\max\limits_i \mathbf{m}[i]\). 
Since the alphabet of the matrix $\mathbf{M}$ has size 2, the vector-matrix multiplication step can be accelerated by the Mailman algorithm to reduce runtime in the same way as in our DMC vector–matrix formulation. The complexity of constructing the matrix $\mathbf{M}$ is amortized. The computational complexity is therefore dominated by the complexity of the vector-matrix multiplication step:
\[
\Oh(n)+\mathcal{O}\left( \frac{n \cdot S}{\log\left(\max\{n, S\}\right)} \right) = \mathcal{O}(S).
\]

This approach can be generalized to \(q\)-ary erasure channels with a similar encoding strategy.

\subsection{Application to Syndrome Decoding for Linear Codes}
\label{subsec:linear_syndrome}

For the special case of a binary linear code over a BSC, ML decoding is equivalent to finding the most likely error pattern. 
Let $\mathscr{C}[n,k]_2$ be a binary linear code with parity-check matrix $\mathbf{H}$, and let 
$\mathcal{L} = \{\boldsymbol{e}_1, \boldsymbol{e}_2, \dots, \boldsymbol{e}_{2^{n-k}}\}$ denote the set of all $2^{n-k}$ distinct coset leaders. 
Upon receiving a vector $\boldsymbol{y} \in \mathbb{F}_2^n$, the decoder’s task is to identify the coset leader $\boldsymbol{e}_j \in \mathcal{L}$ whose syndrome $\boldsymbol{b}_j = \mathbf{H}\boldsymbol{e}_j^\top$ exactly matches the received syndrome $\boldsymbol{s} = \mathbf{H}\boldsymbol{y}^\top$. 
We next show how our framework can be adapted to solve this exact-match problem efficiently, reducing the decoding complexity to $\mathcal{O}\!\left(2^{\min\{k,\, n-k\}}\right)$.

The objective is to compute a vector \(\boldsymbol{d}\) of length \(2^{n-k}\) consisting of Hamming distances, where each entry \(\boldsymbol{d}[j]\) is the distance \(d_H(\boldsymbol{s}, \boldsymbol{b}_j)\) between the received syndrome and the syndrome of the \(j\)-th coset leader. We then find the index \(j^*\) where this distance is zero. To express this computation as a vector-matrix product, we define a specific encoding for the syndrome bits. For any bits \(a, b \in \{0,1\}\), let
\[
v(a) = [1-a, a] \quad \text{and} \quad u(b) = [b, 1-b].
\]
This encoding has the crucial property that the inner product \(v(a) \cdot u(b)^\top = (1-a)b + a(1-b) = \mathds{1}\{a \neq b\}\), which is 1 if the bits differ and 0 otherwise. We now construct the vector and matrix for our syndrome-decoding problem, analogous to \(V(\boldsymbol{y})\) and \(\mathbf{M}\) in the main formulation.

\begin{itemize}
    \item \textbf{Syndrome Vector \(V_{\text{synd}}(\boldsymbol{s})\):} Given the received syndrome \(\boldsymbol{s} \in \mathbb{F}_2^{n-k}\), we construct a real-valued vector of length \(2(n-k)\) by concatenating the \(v(\cdot)\) encodings of its bits:
    \[
    V_{\text{synd}}(\boldsymbol{s}) = [v(s_1), v(s_2), \dots, v(s_{n-k})].
    \]
    \item \textbf{Syndrome Matrix \(\mathbf{M}_{\text{synd}}\):} During a one-time preprocessing, we compute the syndrome \(\boldsymbol{b}_j\) for every coset leader \(\boldsymbol{e}_j \in \mathcal{L}\). We then construct a binary matrix \(\mathbf{M}_{\text{synd}} \in \{0,1\}^{2(n-k) \times 2^{n-k}}\) where each column \(j\) is the concatenated \(u(\cdot)\) encodings of the bits of \(\boldsymbol{b}_j\):
    \[
    \mathbf{M}_{\text{synd}} = \big[ U(\boldsymbol{b}_1)^\top \;\; U(\boldsymbol{b}_2)^\top \;\; \cdots \;\; U(\boldsymbol{b}_{2^{n-k}})^\top \big],
    \]
    where \[U(\boldsymbol{b}_j) = [u(\boldsymbol{b}_j[1]), u(\boldsymbol{b}_j[2]), \dots, u(\boldsymbol{b}_j[n-k])] \in \mathbb{F}_2^{2(n-k)}. \]
\end{itemize}

By this construction, the vector-matrix product directly yields the vector of Hamming distances:
\begin{align*}
\boldsymbol{d} & = V_{\text{synd}}(\boldsymbol{s}) \cdot \mathbf{M}_{\text{synd}},\\ \quad \text{where } \boldsymbol{d}[j] & = \sum_{i=1}^{n-k} v(s_i) \cdot u(\boldsymbol{b}_j[i])^\top = d_H(\boldsymbol{s}, \boldsymbol{b}_j).
\end{align*}
The ML decoding task reduces to computing \(\boldsymbol{d}\) via the Mailman algorithm and finding the index \(j^* = \arg\min\limits_j \boldsymbol{d}[j]\). The most likely error pattern is \(\boldsymbol{e}_{j^*}\), and the decoded codeword is \(\boldsymbol{c}^* = \boldsymbol{y} \oplus \boldsymbol{e}_{j^*}\).

Since the matrix $\mathbf{M}_{\text{synd}}$ is binary, the vector-matrix multiplication step can be accelerated by the Mailman algorithm. The complexity of constructing the syndrome matrix $\mathbf{M}_{\text{synd}}$ is amortized. The computational complexity is therefore determined by the size of the coset leader search space:
\[
\Oh(2(n-k))+\mathcal{O}\left( \frac{2(n-k) \cdot 2^{n-k}}{\log\left(\max\{2(n-k), 2^{n-k}\}\right)} \right) = \mathcal{O}(2^{n-k}).
\]
Combining this with $\mathcal{O}(2^k)$—the complexity established earlier in the general case of ML decoding of $q$-ary codes over DMC, the overall complexity for ML decoding of linear codes on the BSC via this approach is $\mathcal{O}(2^{\min\{k,\, n-k\}})$. It is straightforward to extend the method to ML decoding of $q$-ary symmetric channels using $q$-ary linear codes, in which case the total complexity is $\Oh(q^{\min\{k,\, n-k\}})$.

\subsection{Extension to Intersymbol-Interference (ISI) Channels}
\label{subsec:ISI}
We now extend our method to an \( L \)-tap intersymbol interference (ISI) channel, where the output \( y_i \) at time \( i \) depends on the current and \( L \) previous inputs, characterized by the transition probability \( P(y_i \mid x_i, x_{i-1}, \dots, x_{i-L}) \). When $L=0$, this corresponds to the DMC considered previously. 

The ML decoding rule for a received sequence \( \boldsymbol{y} \) is to find the codeword \( \boldsymbol{c} \in \mathscr{C} \) that maximizes
\begin{align}
    \boldsymbol{c}^* & = \arg \max_{\boldsymbol{c} \in \mathscr{C}} \prod_{i=1}^n P(y_i \mid c_i, c_{i-1}, \dots, c_{i-L})\notag \\
    & = \arg \max_{\boldsymbol{c} \in \mathscr{C}} \sum_{i=1}^n \log P(y_i \mid c_i, c_{i-1}, \dots, c_{i-L}).
    \label{eq:ML_decoding_isi}
\end{align}
The proposed framework can be directly applied by redefining the fundamental unit of encoding. Instead of considering individual symbols from an alphabet of size \( q \), we consider the state defined by the input tuple \( (x_i, x_{i-1}, \dots, x_{i-L}) \). There are \( q^{L+1} \) such possible tuples.

The proposed framework naturally extends to this setting by redefining the fundamental encoding unit. Instead of treating individual input symbols from an alphabet of size $q$, we consider the composite state formed by each input tuple $(x_i, x_{i-1}, \dots, x_{i-L})$. There are $q^{L+1}$ possible such tuples.

Accordingly, the core vectors introduced in Section~\ref{subsec:discrete} are modified as follows:
\begin{align*}
V(\boldsymbol{y})
  &= \big[\, \boldsymbol{v}_1(\boldsymbol{y}),\, \boldsymbol{v}_2(\boldsymbol{y}),\, \dots,\, \boldsymbol{v}_n(\boldsymbol{y}) \,\big], \\
I(\boldsymbol{c})
  &= \big[\, \boldsymbol{u}_1(\boldsymbol{c}),\, \boldsymbol{u}_2(\boldsymbol{c}),\, \dots,\, \boldsymbol{u}_n(\boldsymbol{c}) \,\big].
\end{align*}

\begin{enumerate}
    \item \textbf{Conditional Probability Vector $V(\boldsymbol{y})$:}  
    This vector lies in $\mathbb{R}^{n q^{L+1}}$. Each subvector $\boldsymbol{v}_i$ has $q^{L+1}$ entries, storing the log-likelihood values $\log P(y_i \mid \cdot)$ for all possible input tuples of length $L{+}1$.

    \item \textbf{Incidence Vector $I(\boldsymbol{c})$:}  
    This vector lies in $\{0,1\}^{n q^{L+1}}$. Each subvector $\boldsymbol{u}_i$ is a one-hot encoding of the specific input tuple $(c_i, c_{i-1}, \dots, c_{i-L})$ observed at position $i$ of the codeword.
\end{enumerate}



With these redefined vectors, the ML decoding process remains identical to the memoryless case: we construct a binary matrix \( \mathbf{M} \in \{0, 1\}^{nq^{L+1} \times S} \) from the incidence vectors of all codewords and compute the score vector \( \mathbf{m} = V(\boldsymbol{y}) \cdot \mathbf{M} \), where $S$ denotes the number of codewords in the codebook. The ML codeword corresponds to the maximum entry in \( \mathbf{m} \).

The computational complexity of the vector-matrix multiplication step becomes
\[
\mathcal{O}\left(\dfrac{nq^{L+1}S}{\log(\max\{nq^{L+1},\, S\})}\right).
\]
For a code with \( S = q^k \) codewords and assuming \( nq^{L+1} \le q^k \), the complexity simplifies to
\[
\mathcal{O}\left(\dfrac{nq^{L+1} \cdot q^k}{k \log q}\right) = \mathcal{O}\left(\dfrac{1}{R} \cdot \dfrac{q^{}}{\log q} \cdot q^{k+L}\right) = \mathcal{O}(q^{k+L}),
\]
where \( R = k/n \) is the code rate. The dominant cost is thus determined by the codebook size and channel memory length. 

Following the same logic, the proposed framework for ISI channels also extends straightforwardly to maximum-likelihood list, soft, and erasure decoding.

\section{Simulation}
\label{sec:Simulation}

We evaluate the performance of the proposed decoder against exhaustive search (ESD) and GRAND~\cite{GRAND:DuffyKLM'19} over a memoryless binary symmetric channel (BSC), where each bit is independently flipped with probability \(p\). The choice of ESD and GRAND is because they are universal decoders, just as the proposed decoder. The BSC can be modeled as $\boldsymbol{y} = \boldsymbol{x} + \boldsymbol{e}$, where $\boldsymbol{x}, \,\boldsymbol{e} \in \mathbb{F}_2^n$ are respectively the transmitted codeword and the error pattern, and addition is modulo $2$. To ensure a fair comparison, no attempt was made to significantly optimize or parallelize any of the decoding algorithms.

The benchmark decoders operate as follows:
\begin{itemize}
    \item ESD: Upon receiving a vector \(\boldsymbol{y}\), the decoder scans the entire codebook \(\mathscr{C}\) and outputs the codeword that minimizes the Hamming distance to \(\boldsymbol{y}\).
    \item GRAND: The decoder generates candidate error patterns \(\boldsymbol{e} \in \{0,1\}^n\) in decreasing order of likelihood. Over a memoryless BSC with \(p<\tfrac12\), this ordering coincides with increasing Hamming weight, with ties broken arbitrarily. Decoding terminates when \(\boldsymbol{y}-\boldsymbol{e}\) is a valid codeword in codebook $\mathscr{C}$, 
    or when a predefined maximum number of guesses, denoted by \texttt{max\_guess}, is reached, in which case a \emph{decoding failure} is declared (i.e., guessing with \emph{abandonment}~\cite{GRAND:DuffyKLM'19,GRAND_abandon_analysis}).
\end{itemize}

All simulations are conducted using binary Reed--Muller codes of order \(r\), denoted by \(\mathrm{RM}(r,m)\), with \(0\le r\le m\). Reed--Muller codes have recently attracted renewed interest due to their capacity-achieving behavior over both the BSC and the BEC~\cite{kudekar2017reed, Reeves, abbesandon}. An \(\mathrm{RM}(r,m)\) code has block length \(n=2^m\) and dimension
\(
    k=\sum_{i=0}^{r}\binom{m}{i},
\)
resulting in rate \(R=k/n\)~\cite{Coding:books/MacWilliamsS77}. For fixed \(m\), increasing \(r\) increases the dimension and hence the rate of the code.

\begin{figure*}[t]
  \centering
  \begin{subfigure}[t]{0.45\textwidth}
    \includegraphics[width=\linewidth]{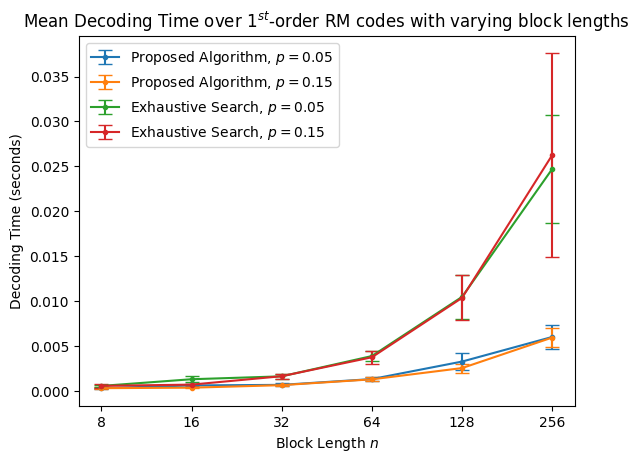}
    \caption{Our proposed method vs. ESD}
    \label{fig:mailman_vs_search}
  \end{subfigure}\hfill
  \begin{subfigure}[t]{0.45\textwidth}
    \includegraphics[width=\linewidth]{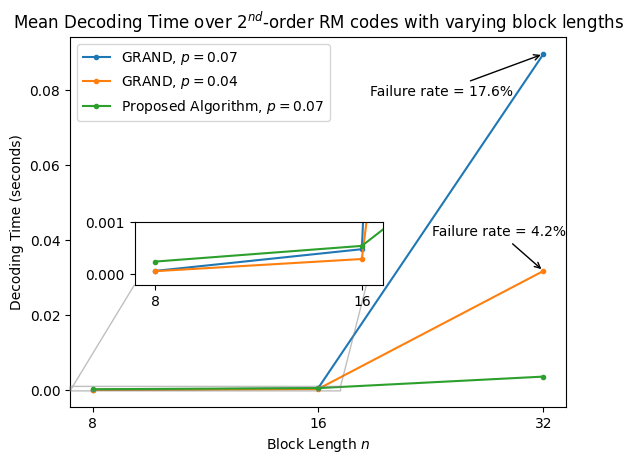}
    \caption{Our proposed method vs. GRAND}
    \label{fig:mailman_vs_grand}
  \end{subfigure}
  \caption{Decoding time performance for our proposed algorithm, ESD, and GRAND in decoding over BSC with various values of bit flip probabilities $p$ and block lengths $n$, using Reed-Muller codes.}
  \label{fig:Simulation}
\end{figure*}

Simulation results are reported for multiple block lengths \(n\) and channel crossover probabilities \(p\). In each experiment, 500 codewords from \(\mathrm{RM}(1,m)\) are randomly generated and transmitted over the BSC. Decoding time is measured from the moment the corrupted word is produced until a (maximum-likelihood) decoded codeword is declared. The reported decoding times correspond to averages over the 500 trials and are summarized in Fig.~\ref{fig:Simulation}.

\paragraph{Comparison with ESD.}
The first plot considers first-order Reed--Muller codes. Since both ESD and the proposed decoder necessarily evaluate all codewords in the codebook, the choice of code rates does not affect the fairness of this comparison. Moreover, for both methods, the bit-flip probability \(p\) does not influence the decoding time, which depends only on the codebook size. At each operating point, we report the mean decoding time together with error bars corresponding to one standard deviation. The proposed decoder exhibits a noticeably smaller variance in running time; for a fixed block length, any residual variation is attributable solely to implementation-level effects.

As shown in Fig.~\ref{fig:mailman_vs_search}, the proposed method consistently outperforms exhaustive search. This behavior is consistent with the theoretical complexity analysis: while exhaustive search has exponential complexity in the parameter \(m\), the proposed decoder achieves an additional speedup by a factor of \(n\). 

\paragraph{Comparison with GRAND.}
The second plot considers $2^{\text{nd}}$-order RM codes with lengths \(n=\{8,16,32\}\), corresponding to moderate-to-high rates (\(R\ge\tfrac12\)), where GRAND is most effective~\cite{GRAND:DuffyKLM'19}. In this regime, \(p\) must remain small, since high-rate codes have lower minimum distance and are not suitable for low-SNR channels~\cite{low_capacity_RM}. For this reason, the largest block length considered is \(n=32\). When \(n=64\) and \(p=0.07\), the expected number of bit flips is \(64 \times 0.07 > 4\). In such cases, GRAND must typically examine at least
\[
    \sum_{i=0}^{4} \binom{64}{i} \approx 6.8 \times 10^{5}
\]
candidate error patterns before identifying the correct one, rendering the algorithm prohibitively inefficient.

The comparison with GRAND is therefore restricted to smaller instances. 
As shown in Fig.~\ref{fig:mailman_vs_grand}, for small values of \(n\), GRAND achieves performance comparable to the proposed decoder, since only a limited number of low-weight error patterns need to be tested. However, when the block length increases to \(n \ge 32\), the rapid growth in the number of  plausible error patterns leads to substantially higher decoding time and an increased failure rate for GRAND. Although increasing \texttt{max\_guess} can reduce failures, it does so at the cost of considerably higher average decoding time.

\section{Conclusion}
\label{sec:Conclusion}

We develop a maximum-likelihood (ML) decoding algorithm for arbitrary block codes by reformulating the ML decoding task as a vector--matrix multiplication problem, after which identifying the ML codeword reduces to selecting the largest entry of the resulting vector. To accelerate this computation, we employ the Mailman algorithm~\cite{mailman_algorithm}, yielding an \(n\)-fold reduction in decoding complexity for a length-\(n\) code. The proposed framework is particularly well suited for latency-critical applications in modern communication systems. Promising directions for future work include developing approximation methods that reduce the associated storage requirements while ensuring near-optimal performance.

Vector–matrix multiplication is a common, basic computational routine in numerical linear algebra, optimization, machine learning, and large-scale graph analytics, among many other applications~\cite{Optimizing_sparse_MaVeD_for_large_data}. Its computational complexity remains a central problem in theoretical computer science, where even mild asymptotic improvements can have a wide impact~\cite{anand2025structural}. For a general $m \times n$ matrix $\mathbf{A}$ (where $m <n$) and an $n$-dimensional vector $\boldsymbol{x}$, the straightforward computation of $\mathbf{A}\boldsymbol{x}$ requires $\Oh(mn)$ time. The key to achieving faster computation lies in exploiting the matrix's discrete structure. Specifically, when $\mathbf{A}$ has entries drawn from a small finite alphabet—as is the case in our construction—the Mailman algorithm can precompute and reuse partial inner products corresponding to recurring column patterns. This method reduces redundant multiplications and yields a logarithmic improvement, lowering the complexity to $\Oh(mn/\log n)$ after a one-time preprocessing step. In our setting, this results in a total complexity of $\Oh(q^k)$, correspond to a $\Omega(n)$ speedup. This complexity of $\Oh(q^k)$ still requires that one examines every possible codeword in the codebook.

Further refinements of the Mailman algorithm improve the complexity of the vector–matrix multiplication routine to $\mathcal{O}(mn / \log^2 n)$~\cite{MaVed_subquadratic_time}. In our setting, this yields an overall runtime of $\mathcal{O}\!\left(\tfrac{q^{k}}{k}\right)$. These finite-alphabet acceleration techniques have also been employed to speed up the decoding of Hidden Markov Models (HMMs), achieving a logarithmic-factor improvement over the Viterbi algorithm~\cite{Decode_HMM}. Beyond such logarithmic gains, however, further improvements appear unlikely: obtaining more than a polylogarithmic speedup over Viterbi has been shown to be theoretically hard~\cite{Viterbi_speedup_hard}, as it would require solving \textsf{NP}-hard problems without exhaustively searching over all possible solutions—an ability that lies at the heart of the \textsf{P} versus \textsf{NP} problem~\cite{SAT_feasibility}.

Moreover, a significant barrier is posed by the widely believed Online Matrix--Vector (OMv) conjecture from theoretical computer science. It asserts that a truly sub-$mn$ time per multiplication—that is, time $o(mn)$ asymptotically smaller than the product of the dimensions—is impossible for general cases. This conjecture is frequently used to establish conditional lower bounds for various computational algorithms across multiple domains~\cite{MaVeD_multiplication_conjecture}. Moreover, additional results show that even with unlimited preprocessing and memory, certain variants of the problem still require near-quadratic time, reinforcing the belief that polynomial improvements are unlikely~\cite{Unconditional_hardness_result}. For this reason, we consider it hard that any exact, universal ML decoding method could achieve substantially lower complexity than our proposed algorithm.

\section*{Acknowledgment}
This was supported in part by  NSF-BSF grant FET-2120262. We thank Swapnil Saha, Mursalin Habib, and Karthik C. S. for valuable input.


\bibliography{bibliography}
\bibliographystyle{IEEEtran}
\appendix
\section*{The Mailman Algorithm for Fast vector-matrix Multiplication}\label{appendix:mailman}
The \emph{Mailman algorithm}, introduced by Liberty and Zucker~\cite{mailman_algorithm}, is a fast vector-matrix multiplication scheme for matrices over small alphabets. It is designed to efficiently compute the product of a matrix \( \mathbf{M} \, \in \, \mathcal{X}^{m \times n} \), where \( \mathcal{X} \subset \mathbb{R} \) is a finite alphabet, and a real-valued vector \( \boldsymbol{x} \, \in \, \mathbb{R}^n \). When the number of rows \( m \) is not too large relative to the number of columns \( n \), and the matrix \( \mathbf{M} \) has repeated column patterns, the algorithm achieves significant speedup over the straightforward \( O(mn) \) complexity by reducing redundancy. This condition makes the method particularly well-suited for structured matrix computing applications, as in distributed computing~\cite{Codes_for_MaVeD}, where $\mathbf{M}$ has a small alphabet but large sizes.

\subsection*{Problem Setup}

Let $\mathbf{M} \, \in \, \mathcal{X}^{m \times n}$ be a matrix with entries from a finite alphabet $\mathcal{X} \subset \mathbb{R}$, and let $\boldsymbol{x} \, \in \, \mathbb{R}^n$ be an input vector. The goal is to compute the product $\mathbf{M}\boldsymbol{x}$ efficiently, especially when $\mathbf{M}$ has low alphabet cardinality (e.g., when $\mathbf{M}$ is binary).

\subsection*{Core Idea for $m = \log_2 n$}

In the special case where $m = \log_2 n$ and $\mathcal{X} = \{0,1\}$, every possible binary column vector of length $m$ appears exactly once in $\mathbf{M}$, or can be indexed via a universal matrix $U_n \, \in \, \{0,1\}^{m \times n}$ whose columns enumerate all pairwise distinct binary strings of length $m$. Any such matrix $\mathbf{M}$ can then be written as
\[
\mathbf{M} = U_n P,
\]
where $P \, \in \, \{0,1\}^{n \times n}$ is a sparse \emph{correspondence matrix} where $P(i,j) = 1$ if the $j$-th column of $\mathbf{M}$ matches the $i$-th column of $U_n$. Then the vector-matrix product reduces to
\[
\mathbf{M} \boldsymbol{x} = U_n (P \boldsymbol{x}).
\]
Since \( P \in \{0,1\}^{n \times n} \) is a sparse permutation matrix containing exactly one nonzero entry per column, computing \( P\boldsymbol{x} \) requires only \( \mathcal{O}(n) \) time. Moreover, its structure allows highly efficient storage: each column can be represented solely by the row index of its nonzero element. Consequently, instead of storing \( n^2 \) binary entries, one needs only \( \mathcal{O}(n) \) integers (or \( \mathcal{O}(n \log n) \) bits), reducing the memory requirement from quadratic to linear in \( n \).

The key remaining step is evaluating $U_n \boldsymbol{z}$ for $\boldsymbol{z} \, \in \, \mathbb{R}^n$.

\subsection*{Recursive Evaluation of $U_n \boldsymbol{z}$}

Let $T(n)$ denote the number of scalar additions and multiplications needed to compute $U_n \boldsymbol{z}$ for $\boldsymbol{z} \, \in \, \mathbb{R}^n$. The recursive structure of $U_n$ is
\[
U_2 = \begin{bmatrix} 0 & 1 \end{bmatrix}, \quad
U_n =
\begin{bmatrix}
\mathbf{0}_{n/2} & \mathbf{1}_{n/2} \\
U_{n/2} & U_{n/2}
\end{bmatrix},
\]
where $\mathbf{0}_{\ell}$ and $\mathbf{1}_{\ell}$ are the all-$0$ and all-$1$ row vectors of length ${\ell}$, respectively.
We thus obtain the following recurrence:
\[
T(n) = T(n/2) + 2n, \quad T(2) = 2 \quad \Rightarrow \quad T(n) \leq 4n.
\]
Hence, $U_n \boldsymbol{z}$ can be computed in $\Oh(n)$ time. Notably, \(U_n \) need not be stored explicitly, since its column \( j \) (\(1 \le j \le n\)) is simply the $m$-bit binary representation of \( j-1 \). As a result, its storage cost is negligible and thus excluded from the total space complexity. The total space complexity is thus $\Oh(n)$, which is the required space to store the binary matrix $P$ of size $n \times n$.

\subsection*{General Case: Arbitrary $m$}

When $m \neq \log_2 n$, we partition $\mathbf{M}$ row-wise into $\lceil m / \log_2 n \rceil$ submatrices, each of height at most $\log_2 n$. That is,
\[
\mathbf{M} =
\begin{bmatrix}
\mathbf{M}^{(1)},\, 
\mathbf{M}^{(2)},\,
\hdots,\, 
\mathbf{M}^{(r)}
\end{bmatrix},
\]
where each $\mathbf{M}^{(i)} \, \in \, \mathcal{X}^{m_i \times n}, \quad m_i \leq \log_2 n$. We apply the Mailman method to each $\mathbf{M}^{(i)} \boldsymbol{x}$ separately, each taking $O(n)$ time, resulting in total complexity of
\[
\Oh\left( \frac{mn}{\log_2 n} \right).
\]
An explicit and convenient upper bound on the total complexity is $\dfrac{4mn}{\log_2 n}$~\cite{Codes_for_MaVeD}.



\end{document}